\documentclass[preprint]{aastex}
\usepackage{graphicx}
\usepackage{psfig}
\usepackage{natbib}
\usepackage[english]{babel}  
\usepackage{subfigure}
\usepackage{url}
\newcommand{\new}{ }
\newcommand{\aftr}{ }

\begin{document}
   \title{3D MHD modeling of twisted coronal loops}

\author{F. Reale\altaffilmark{1,2}, S. Orlando\altaffilmark{2}, M. Guarrasi\altaffilmark{3}, A. Mignone\altaffilmark{4}, G. Peres\altaffilmark{1,2},  A. W. Hood\altaffilmark{5}, E. R. Priest\altaffilmark{5} }

 \altaffiltext{1}{Dipartimento di Fisica \& Chimica, Universit\`a di Palermo,
              Piazza del Parlamento 1, 90134 Palermo, Italy
              E-mail: fabio.reale@unipa.it }
\altaffiltext{2}{INAF-Osservatorio Astronomico di Palermo, Piazza del Parlamento 1, 90134 Palermo, Italy}
\altaffiltext{3}{CINECA - Interuniversity consortium, via Magnanelli 6/3, 40033, Casalecchio di Reno, Bologna, Italy}
\altaffiltext{4}{Dipartimento di Fisica Generale, Universit\`a di Torino, via Pietro Giuria 1, 10125, Torino, Italy}
\altaffiltext{5}{School of Mathematics and Statistics, University of St Andrews, St Andrews, KY16 9SS, UK}

            

 
\begin{abstract}  
   {}
   {We perform MHD modeling of a single bright coronal loop to include the interaction with a non-uniform magnetic field. The field is stressed by random footpoint rotation in the central region and its energy is dissipated into heating by growing currents through  anomalous magnetic diffusivity that switches on in the corona above a current density threshold. }
   {We model an entire single magnetic flux tube, in the solar atmosphere extending from the high-$\beta$ chromosphere to the low-$\beta$ corona through the steep transition region. The magnetic field expands from the chromosphere to the corona. 
   The maximum resolution is $\sim 30$ km.  }
   {We obtain an overall evolution typical of loop models and realistic loop emission in the EUV and X-ray bands. The plasma confined in the flux tube is heated to active region temperatures ($\sim 3$ MK) after $\sim 2/3$ hr. Upflows from the chromosphere up to $\sim 100$ km/s fill the core of the flux tube to densities above $10^9$ cm$^{-3}$. More heating is released in the low corona than the high corona and is finely structured both in space and time. }
\end{abstract}


\section{Introduction}
\label{sec:intro}

Coronal loops are magnetic flux tubes where million degree plasma is confined and are the building blocks of the magnetically closed part of the solar corona. Understanding them means understanding how the corona is structured and powered \citep[see][for a review]{Reale2014a}. Each coronal loop is known to evolve fairly independently of nearby ones, because the major mass and energy transport processes occur only along the magnetic field lines. This remains true when a coronal loop is modelled as a bundle of thinner fibrils. Although the fibrils show overall a collective behaviour, each of them is thermally isolated from the others. On this basis, coronal loops have been largely investigated as single isolated systems by means of one-dimensional models, where the main role of the magnetic field is to guide the mass and energy transport. This approach has been successful in describing the basic physical processes and many features observed in loops \citep[e.g.,][]{Priest1978a,Rosner1978a,Hood1979a,Nagai1980a,Peres1982a,Doschek1982a,Nagai1984a,Fisher1985c,MacNeice1986a,Hansteen1993a,Priest1998a,Antiochos1999a,Reale2000a,Muller2003a,Bradshaw2003a,Cargill2004a,Bradshaw2006a,Reale2008a,Guarrasi2010a,Reale2012b}.

Triggering a loop brightening inside these one-dimensional single-loop models is by an energy input inside a tenuous and cool initial coronal atmosphere. The heating makes the temperature increase rapidly all over the loop, because of the efficient thermal conduction, and drives a strong overpressure rapidly down to the dense chromosphere. The chromosphere expands upwards and fills the coronal part of the loop with hot denser plasma, so that the loop brightens. The following evolution depends on the duration of the heat release. Continuous heating allows the loop to reach quasi-equilibrium conditions at the highest possible density. With a short heat pulse the plasma cooling becomes important: after the heating, the temperature decreases rapidly by the very efficient conduction, but the density decreases much more slowly, leading to an overdensity over most of the loop's life.
The loops are observed to be bright on time scales longer than the cooling times \citep[e.g.,][]{Rosner1978a}; the question is whether the heat release is really gradual and long-lasting, or is instead made of a sequence of short and localised heat pulses distributed in the loop cross-section\citep{Klimchuk2006a,Reale2014a}. In the latter case, the real structure of a loop would be that of a bundle of thinner flux tubes whose thickness is determined by the transverse size of the heat pulse. Evidence for overdensity \citep[e.g.,][]{Lenz1999a,Winebarger2003b}, multi-thermal plasma distribution \citep[e.g.,][]{Warren2011a} and some very hot plasma \citep[e.g.,][]{Reale2009b,Testa2012c,Miceli2012a} supports an impulsive heat release, and the question is now turning to how the energy is stored and released, what is the frequency of the pulses, what is the charging mechanism, what is the local conversion mechanism, and whether by the dissipation of waves or by resistive reconnection. \aftr{Intermittent heating and/or fine structuring is also predicted and discussed by some wave dissipation models, either Alfven \citep{van-Ballegooijen2011a,Asgari-Targhi2012a,Asgari-Targhi2013a,van-Ballegooijen2014a,Cranmer2015a} or kink modes \citep{Antolin2014a,Magyar2016a}.}


A second complementary two-dimensional or three-dimensional approach investigates the way the freed magnetic energy powers coronal flux tubes. The stress of a magnetic flux tube has been studied due to twisting \citep[e.g.,][]{{Rosner1978b},{Golub1980a},Klimchuk2000b,Baty2000a,Torok2003a} and braiding of the field lines \citep{Lopez-Fuentes2010a,Wilmot-Smith2011a,Bingert2011a}. Most efforts have been devoted to study the conditions and effects of the resulting kink instability \citep{Hood1979b,Zaidman1989a,Velli1990a,Baty2000a,Gerrard2001a,Torok2004a}, and to the resulting formation of current sheets \citep{Velli1997a,Kliem2004a} and relaxation due to several  dissipation mechanisms \citep{Hood2009b,Bareford2013a}. MHD simulations have shown the possible importance of local instabilities in the coronal magnetic field to trigger cascades to large-scale energy release \citep{Hood2016a}.

A third approach is a large-scale one that ranges from the low chromosphere to the corona. It takes the magnetograms and the observations of photospheric granules, and of their dynamics, as boundary conditions to determine the structure and evolution of the upper atmosphere. This approach is able to describe the formation and powering of coronal loops in a qualitative or semi-quantitative way \citep{Gudiksen2005a,Bingert2011a}, including the emergence of flux tubes by magnetic twisting \citep{Martinez-Sykora2008a,Martinez-Sykora2009a}, so as to reproduce several observed features, such as a constant cross-section \citep{Peter2012a}, and to help interpret and use data analysis tools \citep{Testa2012b}. Recent work has supported episodic and structured heating due to the fragmentation of current sheets and/or turbulent cascades \citep{Hansteen2015a,Dahlburg2016a}.

\new{Evidence for coherent widespread twisting of magnetic flux tubes has been found on the solar disk \citep{{Wedemeyer-Bohm2012a},De-Pontieu2014a} and well-studied \citep{Levens2015a} from optical and UV observations. 
This represents a natural stressing mechanism of the magnetic field that eventually leads to a relaxation and a release of magnetic energy \citep[e.g.,][]{{Rosner1978b},{Golub1980a},Klimchuk2000b,Lopez-Fuentes2003a}.

\new{Here we set up a numerical experiment for a coronal magnetic flux tube that is anchored in the chromosphere and progressively twisted by the rotation of the plasma at the footpoints. Our approach is a step forward from 1-D loop modeling to allow an active role for the magnetic field, including the expansion of the field lines in the transition region and the energy production from field dissipation.  Our choice has been to assume a relatively simple setup but still including many ingredients of a loop heated by magnetic dissipation. A coherent rotation of the footpoints with some moderate perturbation allows us to have some control on the effects in such a complex MHD system. As new achievements with respect to other previous MHD modeling of twisted loops, our modeling includes a highly non-uniform solar atmosphere and magnetic field with the boundary in the chromosphere. A fundamental target of our work is to reproduce the full typical evolution of a coronal loop, including the chromospheric evaporation driven by magnetic heating excess. This is not an easy task because it requires high spatial resolution \citep{Bradshaw2013a} in a 3D MHD framework. Our model aims also at accurately describing the temperature stratification and evolution to synthesize observables for diagnostics and direct comparison with observations.}

We consider a complete loop atmosphere with a corona connected to two thick chromospheric layers by thin transition regions, immersed in a magnetic field. The magnetic field is arranged to be mostly uniform in the corona and strongly tapering in the chromosphere, where the ratio of  thermal  to  magnetic pressure switches from low to high values ($\beta > 1$). Therefore, our model accounts for the interaction with the magnetic field including the critical region where $\beta$ changes regime. \new{The model also includes heating mechanisms that derive from the dissipation of the magnetic field. The heating is basically determined by the anomalous diffusivity that reconnects the magnetic field above a current density threshold. The currents grow because of the progressive twisting of the magnetic field. The field is twisted by random rotational plasma motions at the loop footpoints, which drag the field. We show that this magnetic stress and dissipation drives a typical coronal loop ignition, structuring and evolution.}

\section{The model}


We consider a box containing a single coronal loop. 
For simplicity of modeling, the loop is then straightened into a magnetic flux tube \aftr{rooted in the photosphere through} two chromospheric layers at opposite sides of the box (top and bottom boundaries) \citep{Guarrasi2014a}. \aftr{These two layers can be treated independently since the loop footpoints are far from each other and therefore located in independent regions of the chromosphere and photosphere.}
\aftr{We consider only the gravity component along the flux tube and, in particular, that of a curved (semicircular) flux tube, i.e., it decreases to zero at the midpoint. This assumption holds as long as the twisted region has a small cross-section with respect to the tube length, as it is in this case.}


Our domain is a 3D
cylindrical box ($r,\phi,z$).  The
box is much broader than the cross-section of the loop. 
The evolution of the plasma and magnetic field in the box is described by solving 
the full time-dependent MHD equations including gravity (for a curved loop), thermal conduction (including the effects of heat flux saturation), radiative losses from optically thin plasma and an anomalous magnetic diffusivity.


The MHD equations are solved in the non-dimensional conservative form:

\begin{equation}
  \label{MHD_mass_cons_eq_1}
  \frac{\partial \rho}{\partial t} + {\bf \nabla} \cdot \left( \rho {\bf u} \right) = 0
\end{equation}

\begin{equation}
  \label{MHD_mom_cons_eq_1}
  \frac{\partial \rho {\bf u}}{\partial t} + {\bf \nabla} \cdot \left( \rho {\bf u}  {\bf u} - {\bf B} {\bf B} + {\bf I} P_{t} \right) = \rho {\bf g}
\end{equation}

\begin{eqnarray}
  \label{MHD_Ene_cons_eq_1}
  \frac{\partial \rho E}{\partial t} + {\bf \nabla} \cdot \left[ {\bf u} \left( \rho E + P_{t} \right)  - {\bf B} \left( {\bf v} \cdot {\bf B} \right) \right] = \nonumber \\
 -{\bf \nabla} \cdot \left[ \left( \eta \cdot {\bf J} \right) \times {\bf B} \right] + \rho {\bf u} \cdot {\bf g} - {\bf \nabla} \cdot {\bf F_{c}} - n_{e} n_{H} \Lambda\left( T\right) + Q
\end{eqnarray}

\begin{equation}
  \label{MHD_Induct_eq_1}
  \frac{\partial {\bf B}}{\partial t} + {\bf \nabla} \cdot \left( {\bf u}  {\bf B} - {\bf B} {\bf u}\right) = - {\bf \nabla} \times \left( \eta \cdot {\bf J} \right)
\end{equation}

 \begin{equation}
 {\bf \nabla} \cdot {\bf B} = 0
 \label{Maxwell_02}
\end{equation}


where:

\begin{equation}
  \label{MHD_P_t}
  P_{t} = p + \frac{{\bf B} \cdot {\bf B}}{2}
\end{equation}

\begin{equation}
  \label{j_curr}
  {\bf J} = \frac{c}{4 \pi} {\bf \nabla} \times {\bf B} 
\end{equation}

\begin{equation}
  \label{MHD_Ene_tot}
  E = \epsilon + \frac{{\bf u} \cdot {\bf u}}{2} + \frac{{\bf B} \cdot {\bf B}}{2 \rho}
\end{equation}

\begin{equation}
  \label{TC_flux_pes_summary}
  {\bf F_{c}} =  \frac{F_{sat}}{F_{sat} + \left| {\bf F_{class}} \right|} {\bf F_{class}}
\end{equation}

\begin{equation}
  \label{TC_flux_class_summary}
  {\bf F_{class}} = k_{||} {\bf \hat{ b}} \left( {\bf \hat{ b}} \cdot {\bf \nabla} T \right) + k_{\bot} \left[  {\bf \nabla} T - {\bf \hat{ b}} \left( {\bf \hat{ b}} \cdot {\bf \nabla} T \right) \right]
\end{equation}

\begin{equation}
  \label{TC_flux_abs_class_summary}
  \left| {\bf F_{class}} \right| = \sqrt{\left( {\bf \hat{ b}} \cdot {\bf \nabla} T\right)^{2}(k_{||}^{2} - k_{\bot}^{2}) + k_{\bot}^{2} {\bf \nabla} T^{2}}
\end{equation}

\begin{equation}
  \label{TC_flux_sat_summary}
 F_{sat} = 5\phi \rho c_{iso}^3
\end{equation}

\noindent are the total pressure ($P_{t}$), induced current density (${\bf J}$), and total energy density $E$ (internal energy $\epsilon$, kinetic energy, and magnetic energy) respectively, $p$ is the thermal pressure, $t$ is the time, $n_{H}, n_e$ are the hydrogen and electron number density, respectively, $\rho = \mu m_{H} n_{H}$ is the mass density, $\mu = 1.265$ is the mean atomic mass \citep[assuming metal abundance of solar values,][]{Anders1989a}, $m_{H}$ is the mass of hydrogen atom,  $ {\bf u}$ is the plasma velocity, ${\bf B}$ is the magnetic field, ${\bf \hat{ b}}$ is the unit vector along the magnetic field, ${\bf g}$ is the gravity acceleration vector for a curved loop, ${\bf I}$ is the identity tensor, $T$ is the temperature, $\eta$ is the magnetic diffusivity, ${\bf F_{c}}$ is the thermal conductive flux (see Eq.~\ref{TC_flux_pes_summary}, ~ \ref{TC_flux_class_summary}, ~ \ref{TC_flux_abs_class_summary}, ~ \ref{TC_flux_sat_summary}), the subscripts $||$ and $\bot$ denote, respectively, the parallel and normal components to the magnetic field, $k_{||} = K_{||} T^{5/2}$ and $k_{\bot} = K_{\bot} \rho^2 / (B^2 T^{1/2})$ are the thermal conduction coefficients along and across the field, $K_{||}=9.2 \times 10^{-7}$ and $K_{\bot} = 5.4 \times 10^{-16}$ (c.g.s. units), $c_{iso}$ is the isothermal sound speed, $\phi = 1$ is a free parameter, and $F_{sat}$ is the maximum flux magnitude in the direction of ${\bf F_{c}}$. $\Lambda\left( T \right)$ represents the optically thin radiative losses per unit emission measure derived from CHIANTI v.$~7.0$ database \citep[e.g.,][]{Dere1997a,Reale2012a,Landi2013b} assuming coronal element abundances \citep{Feldman1992b}. $Q=4.2\times 10^{-5}$ erg cm$^{-3}$ s$^{-1}$ is a volumetric heating rate sufficient to sustain a static corona with an apex temperature of about $8\times 10^5$ K, namely a background atmosphere adopted as initial conditions, \aftr{according to the hydrostatic loop model by \cite{Serio1981a}, \citep[see also][]{Guarrasi2014a}. An estimate of this heating rate can be derived from loop scaling laws \citep{Rosner1978a,Reale2014a} that can be rearranged into $Q \sim 10^{-3} T_6^{3.5} L_9^{-2}$, where $T_6$ and $L_9$ are the temperature and  the loop half-length in units of $10^6$ K and $10^9$ cm, respectively.  }


This heating rate is much lower than the one produced by coronal twisting. We use the ideal gas law, $p = (\gamma -1) \rho \epsilon$. We assume negligible viscosity, except for that intrinsic in the numerical scheme.
%


The calculations are performed using the PLUTO code \citep{Mignone2007a,Mignone2012a}, a modular, Godunov-type code
for astrophysical plasmas. The code provides a multiphysics, algorithmic modular environment particularly oriented toward the treatment of astrophysical flows in the presence of discontinuities of the kind present in the case treated here. The code is designed to make efficient use of massive parallel computers using the message-passing interface (MPI) library for interprocessor communications. The MHD equations are
solved using the MHD module available in PLUTO, configured to compute intercell fluxes with the Harten-Lax-Van Leer approximate Riemann solver, while second order in time is achieved using a Runge-Kutta scheme. A Van Leer limiter for the primitive variables is used. The evolution of the magnetic field is carried out adopting the eight wave formulation introduced by \cite{Powell1999a},
that maintains the solenoidal condition $({\bf \nabla} \cdot {\bf B} = 0)$ at the truncation level.

PLUTO includes optically thin radiative losses in a fractional step
formalism \citep{Mignone2007a}, which preserves the $2^{nd}$-order time
accuracy, since the advection and source steps are at least $2^{nd}$
order accurate; the radiative loss $\Lambda$ values are computed at
the temperature of interest using a table lookup/interpolation
method. The thermal conduction is treated separately from advection
terms through operator splitting. In particular we adopted the
super-time-stepping technique \citep{Alexiades1996a} which has
been proved to be very effective to speed up explicit time-stepping
schemes for parabolic problems. This approach is crucial when high
values of plasma temperature are reached (as during flares), the
explicit scheme being subject to a rather restrictive stability
condition (i.e. $\Delta t \le (\Delta x )^{2} / 2 \eta $ where
$\eta$ is the maximum diffusion coefficient), since the thermal
conduction timescale $\tau_{cond}$ is typically shorter than the dynamical one
$\tau_{dyn}$ \citep[e.g.,][]{Orlando2005a,Orlando2008a}.

Our main simulations required about 30 million hours on 32000 cores of the CINECA/FERMI Blue-Gene high performance computing system.

\subsection{The loop setup}
\label{sec:setup}

We describe a box that contains a typical active region loop, with total length of the coronal section $2L = 5 \times 10^{9}$~cm, which is driven to a temperature  $T \sim 3 \times 10^{6}$~K. The loop atmosphere includes a coronal part that is connected to the chromosphere through a steep transition region. In our configuration the corona is in between two independent chromospheric layers, at opposite sides of the geometric domain. Initially, the loop is relatively tenuous and cool: its
atmosphere is plane-parallel and hydrostatic \citep{Serio1981a} with an apex temperature
about $8 \times 10^{5}~K$. In the transition region the temperature drops to $10^4$ K in less than $10^8$ cm. The temperature is uniform at $10^4$ K in the chromosphere. The density correspondingly increases from $\sim 10^{8}$~cm$^{-3}$ in the corona to $\sim 10^{11}$~cm$^{-3}$ in the upper chromosphere and $\sim 10^{14}$~cm$^{-3}$ in the lower chromosphere.

We consider a magnetic field that expands upwards along the loop because of the change of $\beta$ regime from the chromosphere to the corona. To obtain this as initial condition for our modeling, we follow the same procedure as in \cite{Guarrasi2014a}, i.e., we carry out a preliminary 2.5D simulation in cylindrical geometry \aftr{that starts from a magnetic field with field lines running parallel along the loop. This initial magnetic field links the two chromospheres, but its intensity and the background pressure are more intense in the central part of the loop (around the symmetry axis) than in the surroundings. In this simulation, we let this system evolve, and it relaxes to a new equilibrium:} the magnetic field expands considerably in the corona, because the internal total pressure is higher than outside, until the system is again in equilibrium, \aftr{i.e., the maximum plasma velocities are not larger than a few km/s everywhere in the domain}. Finally, we map the 2.5D simulation output into the 3D domain.


\aftr{At the end of this preliminary step the loop is in a new equilibrium: the plasma stratification is very similar to the previous hydrostatic one, but now the magnetic field lines expand from chromosphere
to the corona around the loop central axis}. The magnetic field intensity decreases from  \aftr{$\sim 300$ G at the bottom of the chromosphere to $\sim 60$ G in the transition region and to $\sim 12$ G (still sufficient to confine the loop plasma) at the top of loop, i.e., in the middle of the domain. The area expands} by a factor of 6 from the bottom of the chromosphere to the top of the transition region, another factor 2 in the first 3000 km above the transition region, and a further factor 2 to the top of the loop (middle of the domain). Fig.~\ref{fig:ini} shows the \aftr{equilibrium} conditions from which we start the loop twisting. In this and the following figures the loop is presented as ``straightened'' with the two footpoints and chromospheric layers at the top and bottom of the figure, and the coronal loop in between. The magnetic field is more intense around the central axis and the atmosphere readjusts there to a slightly higher coronal temperature and lower density \aftr{than those reported at the beginning of this section, (the initial atmosphere around the central axis is shown as black lines in Fig.~\ref{fig:prof})}. The field lines clearly show the tapering from the corona to the chromosphere.

The computational domain is 3D cylindrical ($r,\phi,z$, Fig.$~\ref{fig:ini}$). In order to obtain a good compromise between adequate resolution and reasonable coverage of the azimuthal ($\phi$) domain we model only one quadrant with periodic boundary conditions. We also skip the singular central axis, and consider an inner boundary radius $r_{0} > 0$. In the end, the domain range is $-z_M < z < z_M$ along the loop axis where $z_M = 3.1 \times 10^{9} $~cm, $r_0=7 \times 10^{7} \leq r \leq r_M=3.5 \times 10^{9}$~cm across the loop, and  $0 \leq  \phi \leq 90^o$ in the azimuthal direction. 

To describe the transition region at sufficiently high resolution \citep{Bradshaw2013a}, the cell size there ($\left| z \right| \approx 2.4 \times 10^{9}$~cm) decreases to $dr \sim dz \sim 3 \times 10^{6}$~cm. The resolution is uniform in the angle $\phi$, i.e. $d \phi \approx 0.35^o$.

\begin{figure}[!ht]               
\centering
  {\includegraphics[width=8cm]{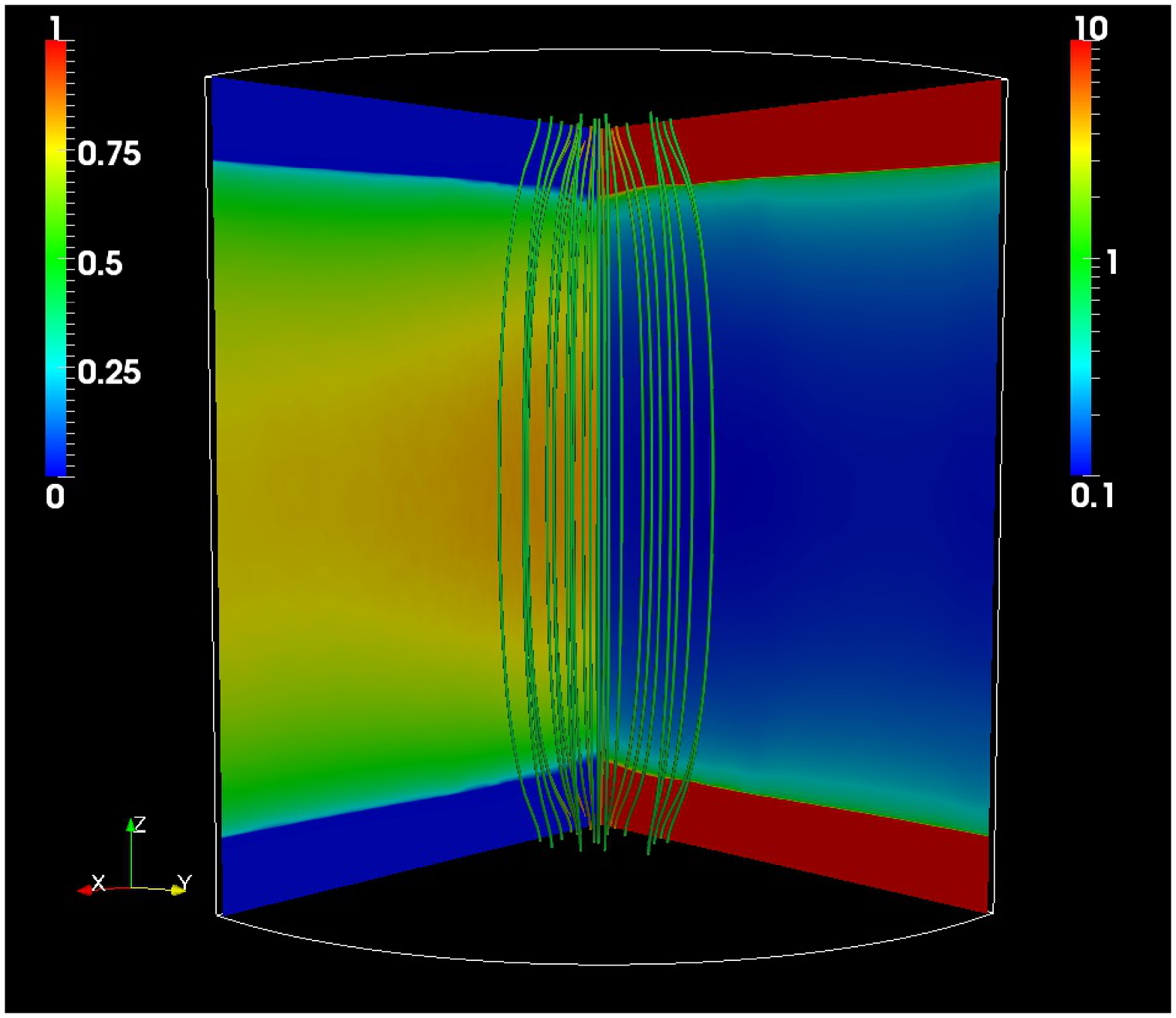}}
\caption{\small Temperature (left, [MK], linear scale) and density (right, [$10^9$ cm$^{-3}$], logarithmic scale) color map in a transverse plane across the loop axis, before the footpoints begin to rotate and twist the magnetic field.  The chromosphere is red in the density map and blue in the temperature map. The magnetic field lines are marked (green lines). }
\label{fig:ini}
\end{figure}

We adopt: reflective boundary conditions at $r = r_0$, i.e. close to the symmetry axis; reflective boundary conditions at $r=r_M$; periodic boundary conditions at $\phi = 0$ and $\phi = 90^o$; and reflective boundary conditions but with reverse sign for the tangential component of the magnetic field  at $z = \pm z_M$. 

\subsection{Loop twisting}

The aim of this work is to trigger heating and brightening of a coronal loop through progressive stressing of the magnetic field twisted at the footpoints. The dissipation is due to anomalous magnetic diffusivity, and grows because the twisting amplifies induced electric currents. The twisting is driven by a rotational plasma motion at both footpoints. \aftr{The driver is photospheric and the rotational motion is set at the lower and upper boundaries of the domain, which can be considered as the boundaries between the photosphere and the chromosphere.} 

The basic rotation profile is that of a rigid body around the central axis, i.e. the angular speed is constant in an inner circle and then decreases linearly in an outer annulus. \aftr{More specifically, at the lower and upper boundaries, the velocity component along $\phi$ is defined as}:

\begin{equation}
v_{\phi} = \pm \omega r \left[1+0.2 \sum_{i = 1}^{6} \sin(\phi\alpha_i) \sin\left(\frac{r}{R_{max}}\alpha_i\right) \right]
\end{equation}

\noindent
\aftr{where the sign is positive (negative) at the lower (upper) boundary,}

\[
\omega = v_{max}/R_{max} \times \left\{
\begin{array}{ll}
\displaystyle 1                     &   \displaystyle r < R_{max} \\  \\
\displaystyle (2 R_{max}-r)/R_{max} &   \displaystyle R_{max} < r < 2R_{max} \\  \\
\displaystyle 0                     &   \displaystyle r > 2R_{max} \\
\end{array}\right.
\]

\noindent
\aftr{The parameters $\alpha_i$ are random numbers between 0 and 30: }

\[
\alpha = [23, 9, 0.6, 15, 19, 13]
\]

\noindent
\aftr{and $v_{max} = 5$~km s$^{-1}$ and $R_{max} = 3000$~km.}


The angular speed $\omega$ was chosen so as the maximum tangential speed is $v(0,R_{rot},t) = 5$ km/s, in agreement with typical photospheric granule speeds \citep{Muller1994a,Berger1996a}. The rotation period of each footpoint is $T_{rot} \approx 1$ hr. The rotational motion is the same but in the opposite direction at the other boundary, i.e. $v(Z_{max},r,t) = -v(0,r,t)$. Since we have equal but opposite motions at the two footpoints, the relative rotation speed of one footpoint with respect to the other is twice, i.e. 10 km/s, and the rotation period is half, i.e. $\sim 1/2$ hr. The input Poynting flux through each rotating footpoint grows linearly to $F_x \sim 3.1 \times 10^7$ erg cm$^{-2}$ s$^{-1}$ at the final time $t \sim 2500$ s. The final total rotation angle is $\approx 2.7 \pi$.

\new{For a more realistic speed pattern, we perturb the velocity through a combination of random sinusoidal functions that depend on $\phi$ and $r$}. The amplitude of these perturbations is 20\%. 
The rotation velocity field is shown in Fig.~\ref{fig:pvel}.}

\begin{figure}[!ht]               
\centering
  {\includegraphics[width=6cm]{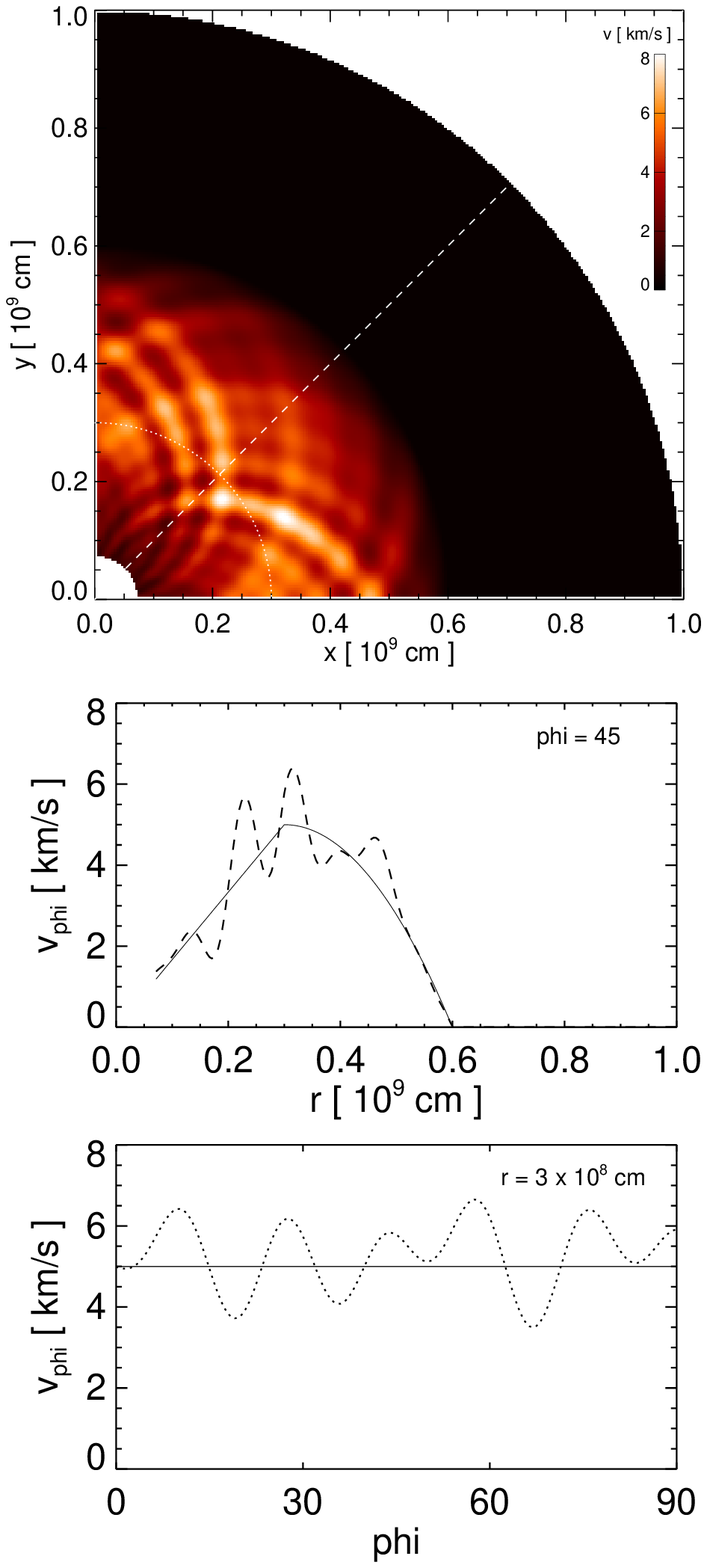}}
\caption{\small \aftr{Rotation velocity field. Top: color map of $v_{\phi}$; middle: velocity profile along $r$ for $\phi =45^o$ (along the dashed line in the top panel) with (dashed line) and without (solid line) random perturbations; bottom: velocity profile along $\phi$ for $r = 3 \times 10^8$ cm (along the dotted line in the top panel) with (dotted line) and without (solid line) random perturbations }.  }
\label{fig:pvel}
\end{figure}

\subsection{The plasma resistivity}
\label{sec:resist}

\new{For our reference simulation of this work we consider an anomalous plasma resistivity that is only switched on when the magnitude of the current exceeds a critical value as in the following \citep{Hood2009b}:}

\begin{equation}
\eta = \left\{ \begin{array}{ll}
\eta_0 & \textrm{$|J| \geq J_{cr}$} \\
0 & \textrm{$|J| < J_{cr}$} 
\end{array} \right\}
\end{equation}

\noindent {where \aftr{we assume $\eta_0 = 10^{14}$ cm$^2$ s$^{-1}$ and $J_{cr} = 75$ A cm$^{-2}$.} 


With this assumption the minimum heating rate above switch on is $H = \eta_0 ( 4 \pi | {J_{cr}} | /c)^2 \approx 0.1$ erg cm$^{-3}$ s$^{-1}$, corresponding to a maximum temperature of $\sim 6$ MK for an equilibrium loop with half length $2.5 \times 10^9$ cm, according to the loop scaling laws \citep{Rosner1978a}. Below the critical current, a minimum numerical resistivity is anyway present, but it does not produce perceptible heating during the simulation. For comparison, we made a simulation also} with a
constant and uniform resistivity in the corona \citep{Bingert2011a}; we set $\eta = 10^{13}$  cm$^2$ s$^{-1}$, which corresponds to a magnetic Reynolds number $R_M \sim 1$ for typical speeds of $\sim 10$ km/s and scale lengths of $\sim 100$ km. In the chromosphere we assume a perpetual perfect equilibrium of energy losses and gains, and therefore we assume a resistivity $\eta = 0$ there. Although the currents are larger in the chromosphere, their dissipation would not increase the chromospheric temperature significantly, because of the very high heat capacity of the dense chromospheric plasma. On the other hand, the current dissipation would weaken considerably the magnetic field, and we have no way to replenish it from below as in the real Sun. Our choice allows us to maintain a sufficiently strong magnetic field throughout the simulation and thus to sustain the coronal heating for a sufficiently long time to reach high temperatures. 

\section{The results}

\new{We model the 3D MHD flux tube (loop) evolution until the loop plasma reaches a maximum temperature $T \sim 4$ MK, i.e. in the time range $0 < t < 2500$ s.}

\begin{figure}[!ht]               
\centering
  {\includegraphics[width=10cm]{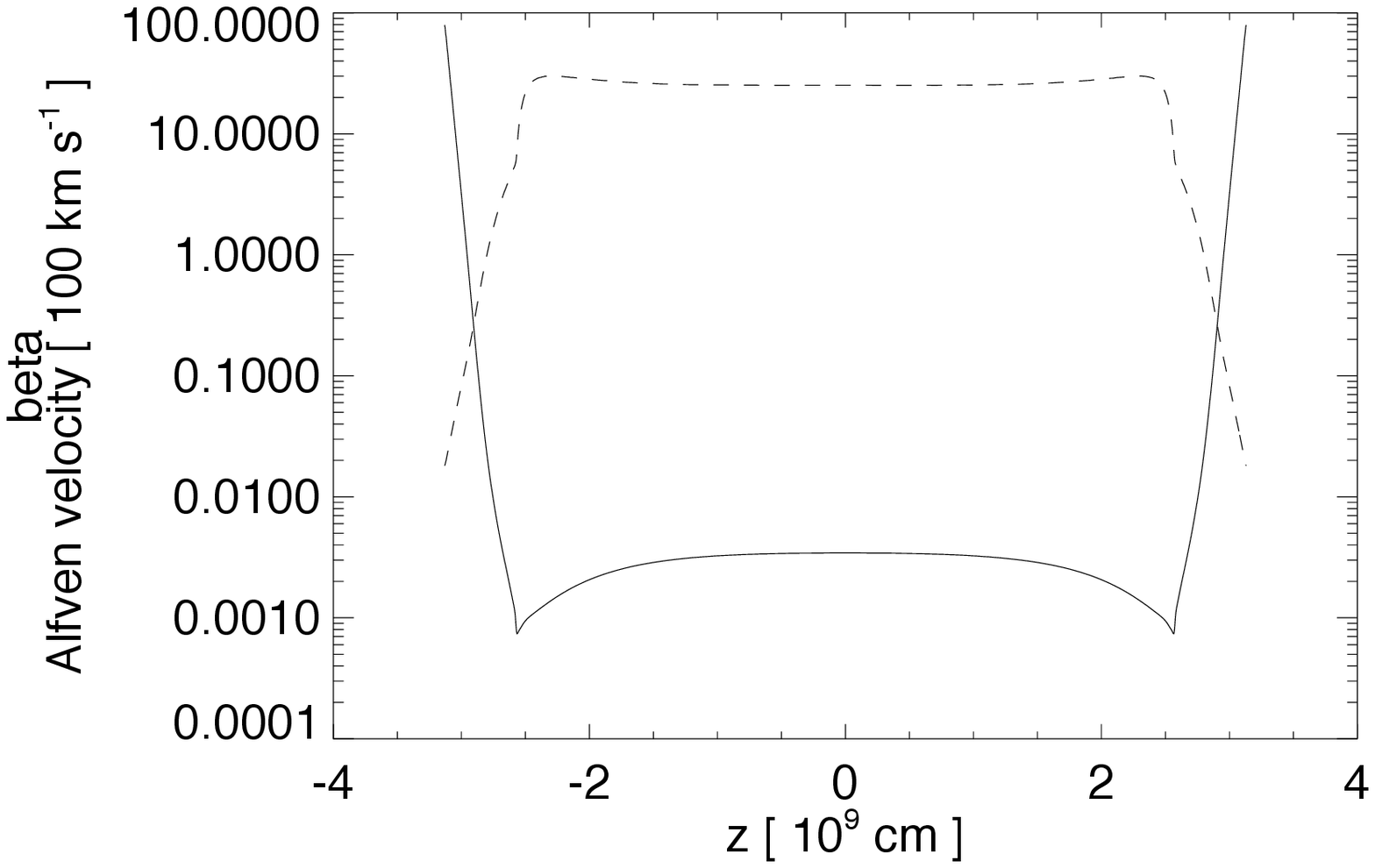}}
\caption{\small Profile of plasma $\beta$ (solid) and Alfven speed (dashed) along the central axis of the domain. }
\label{fig:betalfven}
\end{figure}

The footpoints rotation starts at time $t = 0$. The rotation drags the magnetic field  anchored at the footpoint and the field lines begin to twist since $\beta \sim 100$ there (Fig.~\ref{fig:betalfven}). The twisting propagates upwards at the Alfven speed, whose profile along the flux tube is shown in Fig.~\ref{fig:betalfven}. Below the corona the Alfven speed varies steeply from $\sim 2$ to $\sim 2000$ km/s. The perturbation takes about 200 s to propagate along a vertical distance of $\sim 7000$ km to above the transition region, i.e. with an average speed of $\sim 35$ km/s.

\begin{figure}[!ht]               
\centering
  {\includegraphics[width=12cm]{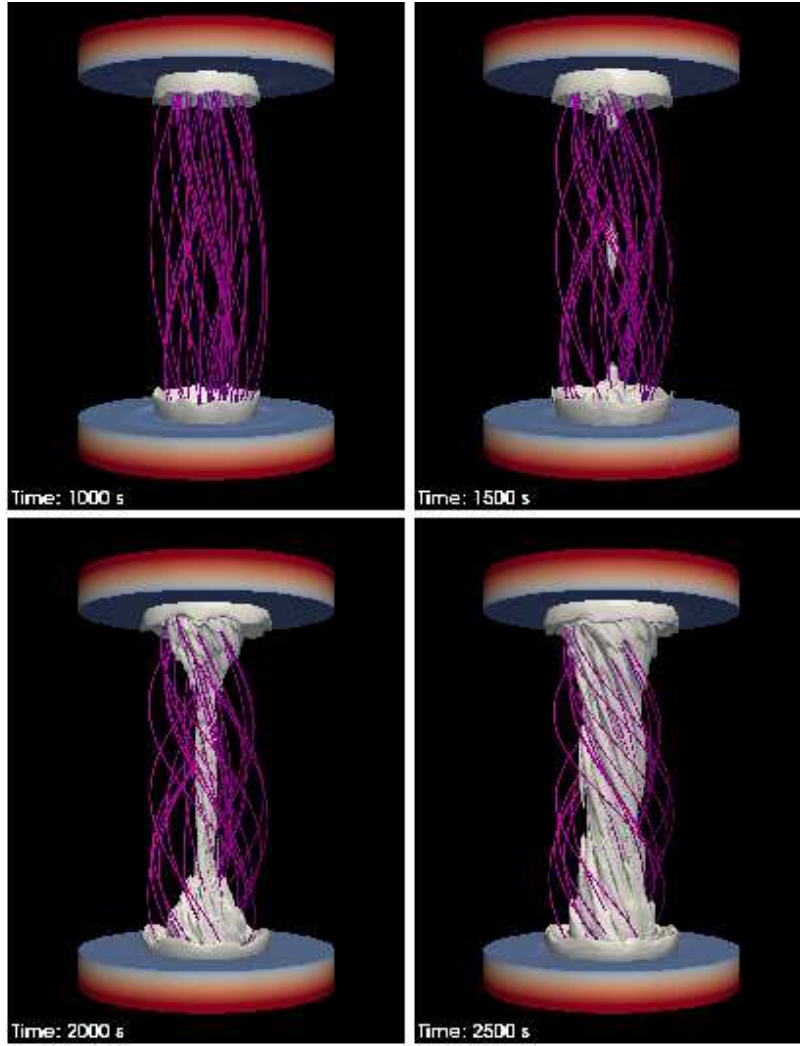}}
\caption{\small Current density surfaces \aftr{(white)} at the critical value \aftr{for dissipation} ($|J_{cr}| = 75$ A cm$^{-2}$) at 4 different times during the twisting of the coronal loop. \aftr{The images result from the replication of the original 90$^o$ domain. Only the region around the central axis is shown. The twisting is around the central vertical axis and} the chromosphere appears as two thick solid \aftr{(colored)} disks at the top and bottom of the domain. Magnetic field lines are also shown \aftr{(pink lines, {\it See on-line movie 1}).} }
\label{fig:evol3d_jcur}
\end{figure}

The progressive twisting of the magnetic field makes the current density gradually increase as well, \aftr{according to Eq.(\ref{j_curr})}. It takes several minutes for the current to grow above the critical value in the corona, which triggers the dissipation into heating. Figure~\ref{fig:evol3d_jcur} shows snapshots of current surfaces at four progressive times during the evolution. The figure shows the current surfaces at the critical value \aftr{for dissipation}. \aftr{Although the computational domain extends over an angle of 90$^o$, for the sake of clarity we replicate the image to cover all 360$^o$.}  The currents are more intense in the low part of the flux tube, where the magnetic field expands and the twisting is driven. The current density first increases in the shell boundary layer of the twisted region (i.e., at $r \sim 6000$ km), where there is a shear between twisted and untwisted region ($t = 1000$ s). Later, the current intensity increases more significantly in the core of the flux tube, as the twisting becomes more and more effective ($t = 1500, 2000$ s). Current intensification propagates from the footpoints upwards all along the flux tube. 


\aftr{As a consequence of the random twisting, the current does not grow uniformly, but it develops into long irregular structures along the field lines, best visible at the final time ($t = 2500$ s). As soon as the current threshold for dissipation is exceeded, the magnetic field lines progressively reconnect in the corona, where the heating is released. 
According to \cite{Schindler1988a}, a signature of the reconnection is the integral of the parallel component of the electric field along the magnetic field lines ($\int E_\parallel dl$), which should be zero without reconnection, since ${\bf E} = {\bf v} \times {\bf B}/c$, where $c$ is the speed of light. Typical values of the electric field can be estimated from the Ohm's law and using the critical current $ |E| = 4 \pi \eta_0 |J_{cr}|/c^2 \sim \pi 10^5$ statV/cm. We find that the integral progressively rises with time from $\sim 10^{13}$ to $\sim 10^{15}$ statV, where the heating is on (and zero elsewhere), confirming substantial reconnection.}



\begin{figure}[!ht]               
\centering
  {\includegraphics[width=12cm]{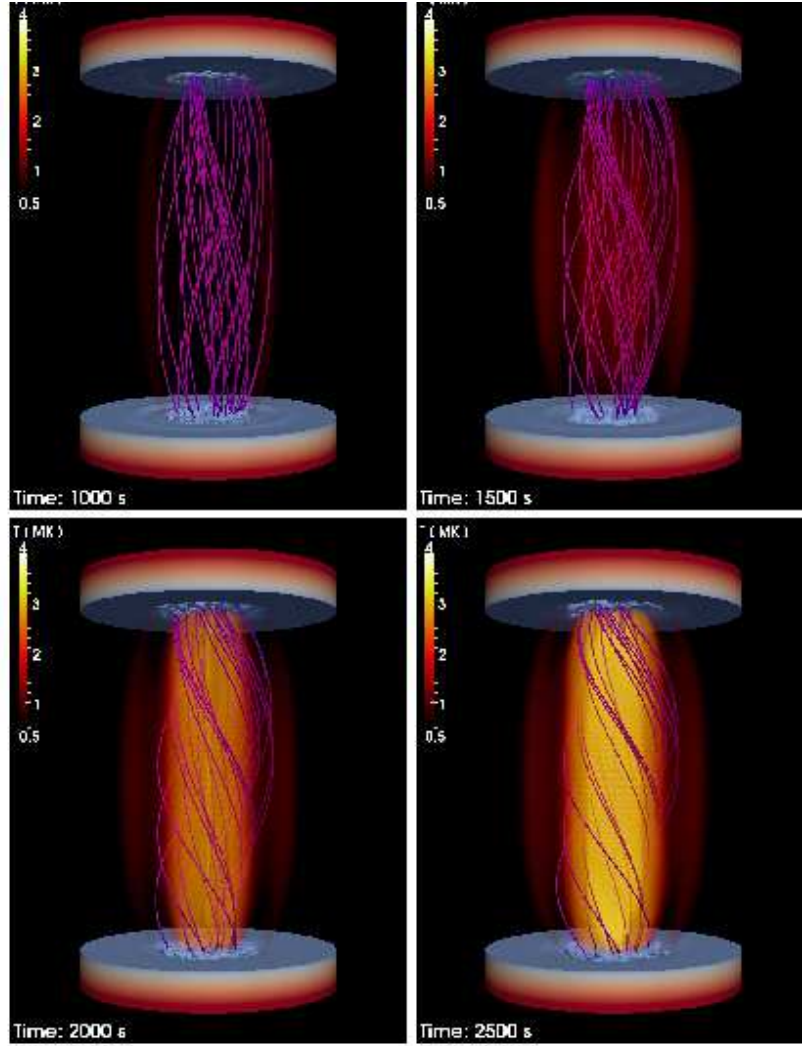}}
\caption{\small \aftr{Temperature rendering at the same times and in the same domain as in Fig.~\ref{fig:evol3d_jcur}.} The units are [$10^6$ K]. Magnetic field lines are also shown (pink lines, {\it See on-line movie 2}). }
\label{fig:evol3d_t}
\end{figure}

\begin{figure}[!ht]               
\centering
  {\includegraphics[width=12cm]{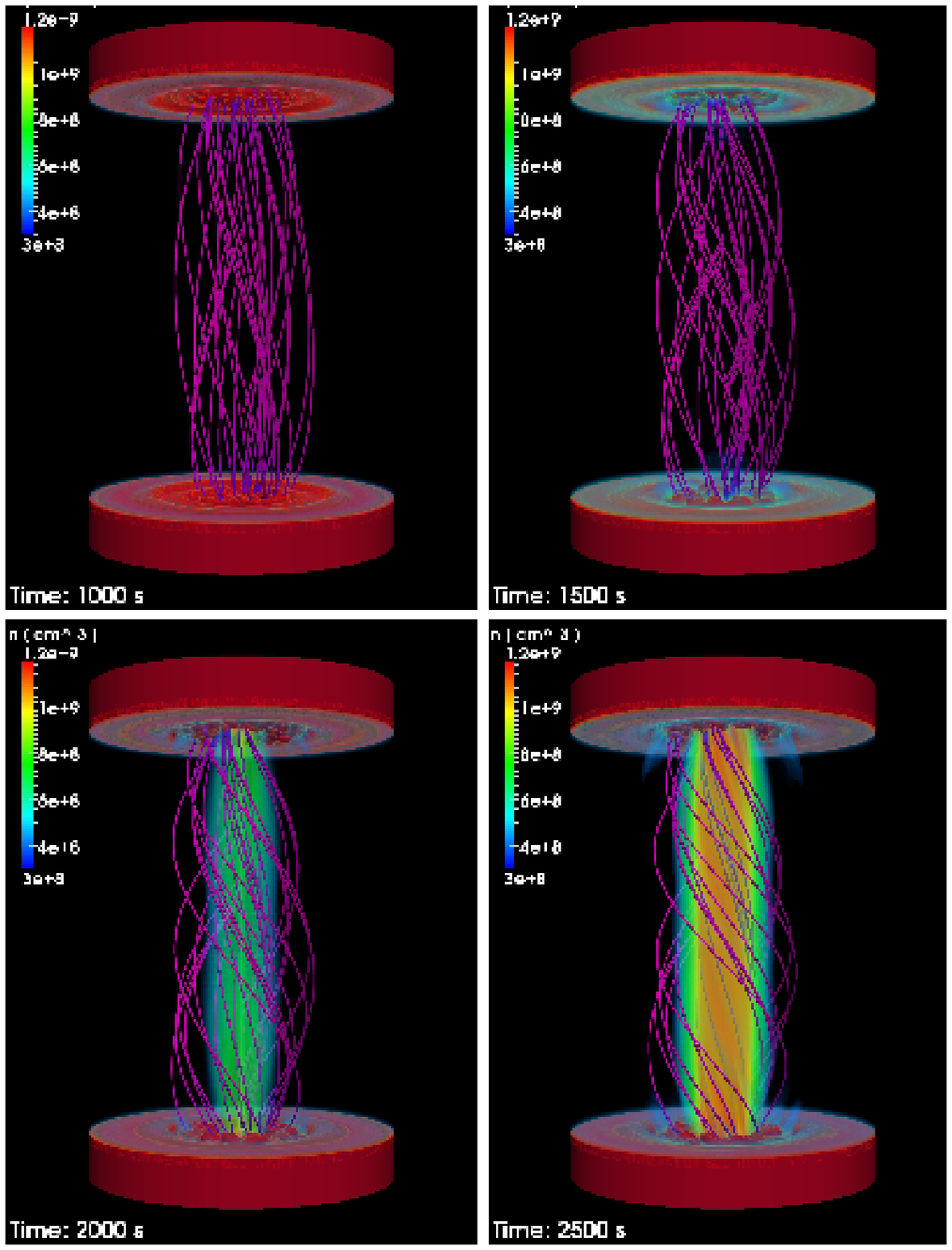}}
\caption{\small \aftr{Density rendering at the same times and in the same domain as in Fig.~\ref{fig:evol3d_jcur}.} The units are [$10^9$ cm$^{-3}$]. Magnetic field lines are also shown (pink lines, {\it See on-line movie 3}). }
\label{fig:evol3d_n}
\end{figure}

Figures~\ref{fig:evol3d_t} and \ref{fig:evol3d_n} show snapshots of the plasma temperature and density at the same times as Fig.~\ref{fig:evol3d_jcur}. The temperature begins to increase significantly at $t \sim 1500$ s and, first, in a shell at the boundary of the twisted region, because of the shear between twisted and untwisted layers. The heating of this outer shell remains quite low throughout the subsequent evolution and the shell is not significantly activated. With some delay, the inner part of the twisted magnetic cylinder is heated and the heating there is more efficient. The inner twisted region becomes hotter quite uniformly all along the flux tube axis, because of the efficient thermal conduction along the field lines in the corona. The evolution of the density is more gradual, i.e. it increases significantly at later times. The density increases because the heating produces an overpressure inside the twisted region, and therefore an expansion of the dense lower layers upwards to the tenuous corona, the so-called {\it chromospheric evaporation}.


\begin{figure}[!ht]               
\centering
  {\includegraphics[width=6cm]{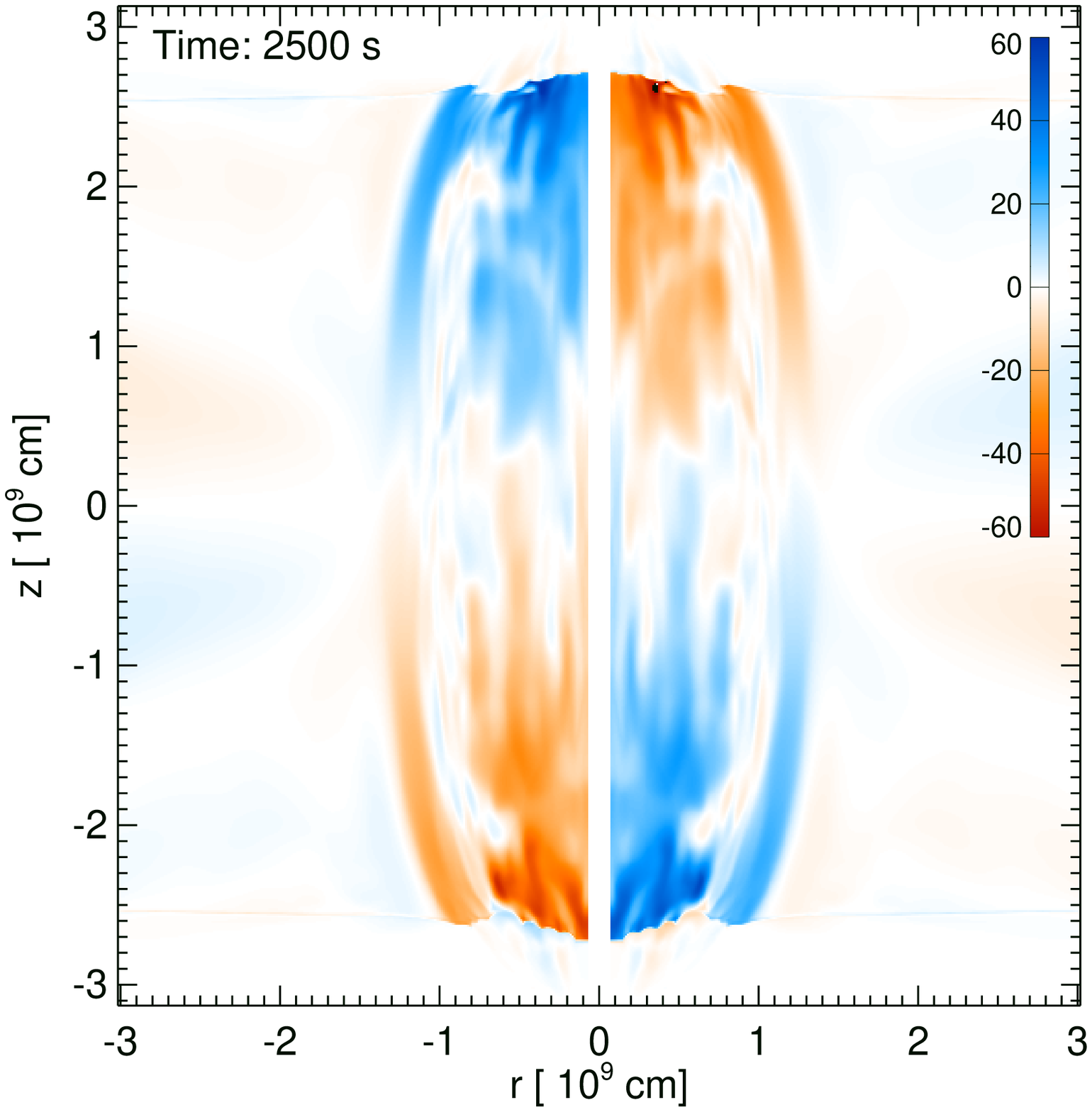}}
\caption{\small Color map of the velocity component [km/s] along $\phi$ in a transverse plane across  the loop axis (see Fig.~\ref{fig:ini}) at time t = 2500 s (blue is outward from the figure, red is inwards).  }
\label{fig:v2d}
\end{figure}

The plasma moves along the magnetic field lines and the twisting of the field makes the motion a spiralling one, adding a significant component along $\phi$. Fig.~\ref{fig:v2d} clearly shows this spiralling component of the upflows from the chromosphere, which would produce blue- and red-shifts if the loop is viewed from the side, as found in recent observations of twisting motions \citep{De-Pontieu2014a}.

The density never grows much above $\sim 3 \times 10^8$ cm$^{-3}$ in the outer shell of the twisted tube. In the core, instead, the coronal density gradually increases to higher values ($\sim 10^9$ cm$^{-3}$), filling the space between the chromospheres. In the end, a proper coronal loop forms, with a dense and hot inner cylindrical region and a thin and more tenuous shell. Looking carefully, especially at the footpoints,  at time $t = 2500$ s, it is possible to distinguish some fine structuring, due to the jagged current dissipation (Fig.~\ref{fig:evol3d_jcur}). The fine structure is less remarkable up in the corona both because of the efficient thermal conduction along the field lines and because of the cross-field dispersal driven by the reconnection \citep{Schrijver2007a}.

\begin{figure}[!ht]               
\centering
  {\includegraphics[width=10cm]{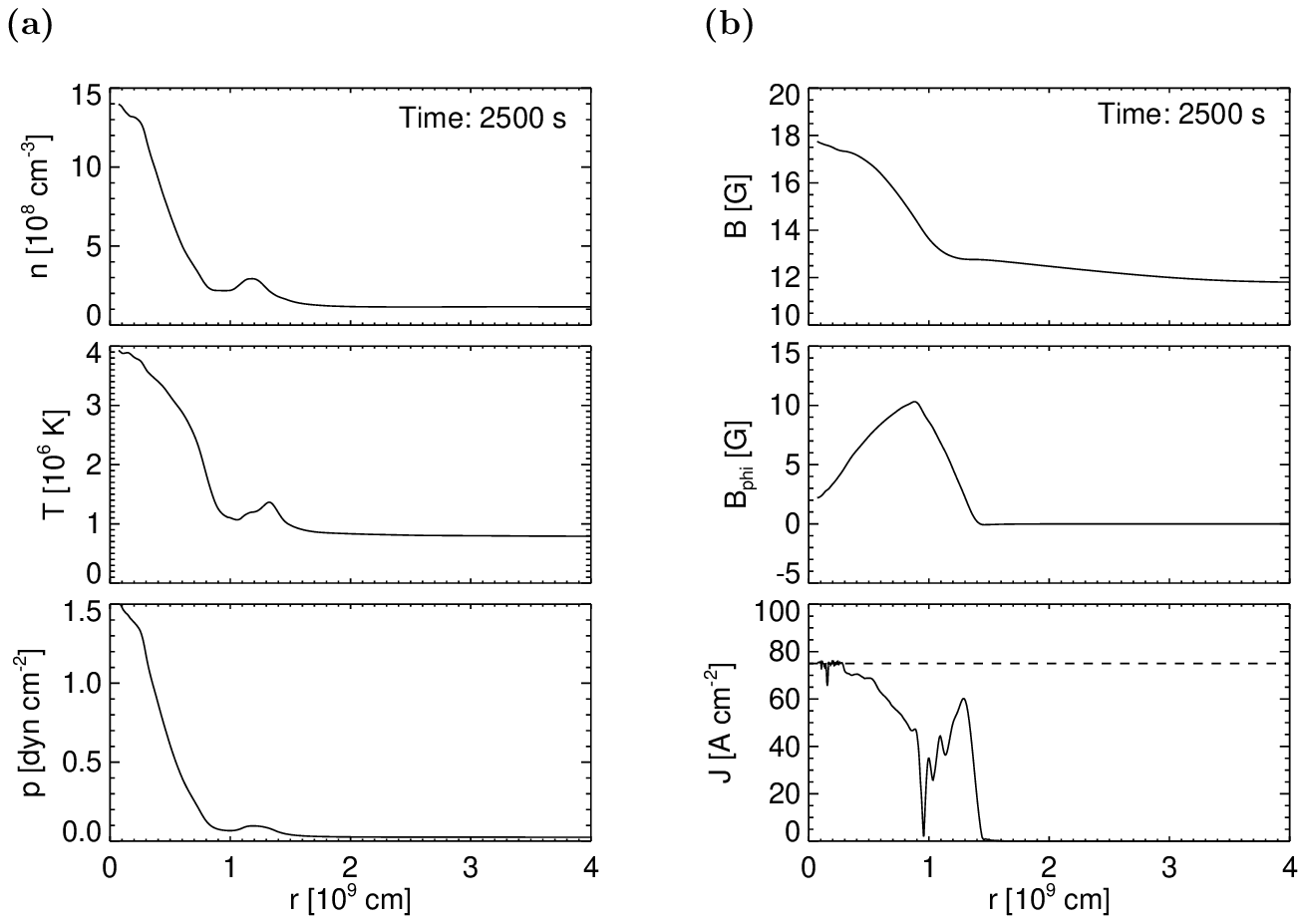}}
  \caption{\small Radial profiles of (a) density, temperature and thermal pressure and (b)   magnetic field intensity, \aftr{azimuthal component of the magnetic field, modulus of the current density, at the top of the loop ($z=0$), for $\phi=45^o$ and at time t = 2500 s.} \aftr{The current threshold for dissipation is marked (dashed line).}  }
\label{fig:radial}
\end{figure}

Fig.~\ref{fig:radial} shows radial profiles of the density, temperature, pressure, magnetic field intensity, \aftr{azimuthal component of the magnetic field}, and current density at the top of the loop and at the end of our simulation. The inner region is the one with the highest values of most quantities, as expected. The first four profiles show a decay from the central axis, to reach a value close to the ambient one at $r > 10^9$ cm. The density decreases more rapidly, because the heating is more effective close to the central axis. The temperature decreases instead more smoothly, as is also perceptible in Fig.~\ref{fig:evol3d_t}. \aftr{The azimuthal component of the magnetic field provides information about the twisting along the loop. The profile at the loop apex is very similar to the unperturbed rotation profile shown in Fig.~\ref{fig:pvel} (middle panel), but widens to a larger radius because of the expansion of the magnetic field.} The current density profile is flat around the critical value for dissipation, $J_{cr}$, to $r \sim 3 \times 10^8$ cm, which is also where the density is the highest.  A secondary peak of $T, p, n$ and $J$ is found at $r \sim 1.2 \times 10^9$ cm, \aftr{and drops} at the boundary between the twisted and untwisted region. The pressure halves at $r \sim 4-5 \times 10^8$ cm, which may be taken as an effective loop width. The magnetic field is amplified by a factor 1.5 around the central axis.

\begin{figure}[!ht]               
\centering
  {\includegraphics[width=10cm]{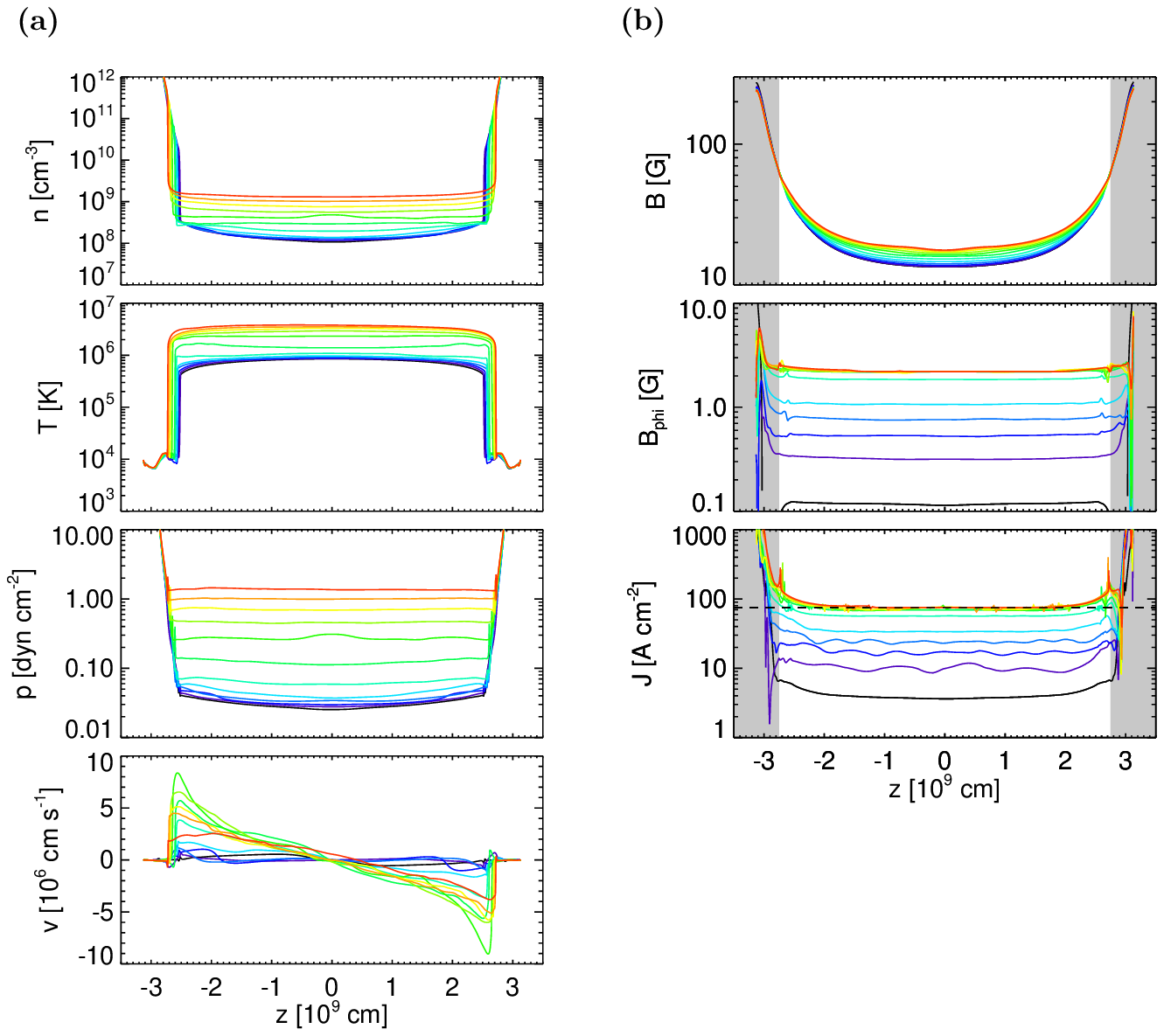}}
\caption{\small Evolution of (a) the plasma density, temperature, thermal pressure and vertical velocity and (b) \aftr{of the total magnetic field intensity, the azimuthal component of the magnetic field,} and the current density, close to the central vertical axis of the twisted flux tube. The profiles are spaced by 200 s and the color coding marks the time progression, from black (t = 0) to red (t = 2500 s). Positive velocity is to the right, i.e. upwards from the left footpoint. The critical current density and the region where the resistivity is zero are marked (horizontal dashed line and grey strips, respectively). }
\label{fig:prof}
\end{figure}

Fig.~\ref{fig:prof} shows profiles of plasma density, temperature, pressure, and vertical velocity \aftr{and of the total magnetic field intensity, the azimuthal component of the magnetic field and the current density} near the central vertical axis of the magnetic flux tube at equi-spaced times. The central axis is a very good approximation of a field line at any time, so we are also looking at the evolution along a field line. From Fig.~\ref{fig:prof}a the loop plasma remains quite steady for a relatively long time at the beginning of the simulation, when the twisting is still unable to provide current dissipation. After several hundreds seconds (pale blue, green lines), the current density increases above the threshold for dissipation in the low corona (i.e. for $|z| \approx 2.6 \times 10^9$ cm), as shown in Fig.~\ref{fig:prof}b, while this occurs earlier at larger radial distances from the central axis. Above this threshold, the heating turns on impulsively, and the density, temperature and pressure all begin to increase rapidly. The profiles are typical of the evolution from standard loop models \citep[e.g.,][]{Reale2000b,Warren2002a,Spadaro2003a,Guarrasi2014a}. The temperature and density appear to increase almost simultaneously, because the evaporation of chromospheric plasma occurs in times smaller than the time spacing of the figure\citep[e.g.,][]{Reale2014a}. The evaporation speeds are higher close to the footpoints. They increase initially up to almost 100 km/s and then begin to settle down to more moderate values below 50 km/s.


Once the heat release has started, the thermal pressure increases regularly and eventually grows above $1$ dyne cm$^{-2}$ in the corona. Fig.~\ref{fig:prof}b shows that \aftr{the magnetic field is progressively twisted, i.e, $B_{\phi}$ increases, uniformly in the corona. It also shows that the total coronal magnetic field increases as well} by about 50 \%, while it does not at the footpoints. The twisting of the magnetic field leads to a boost of the current density in the corona, which is the origin of the heating. 

\begin{figure}[!ht]               
\centering
  {\includegraphics[width=10cm]{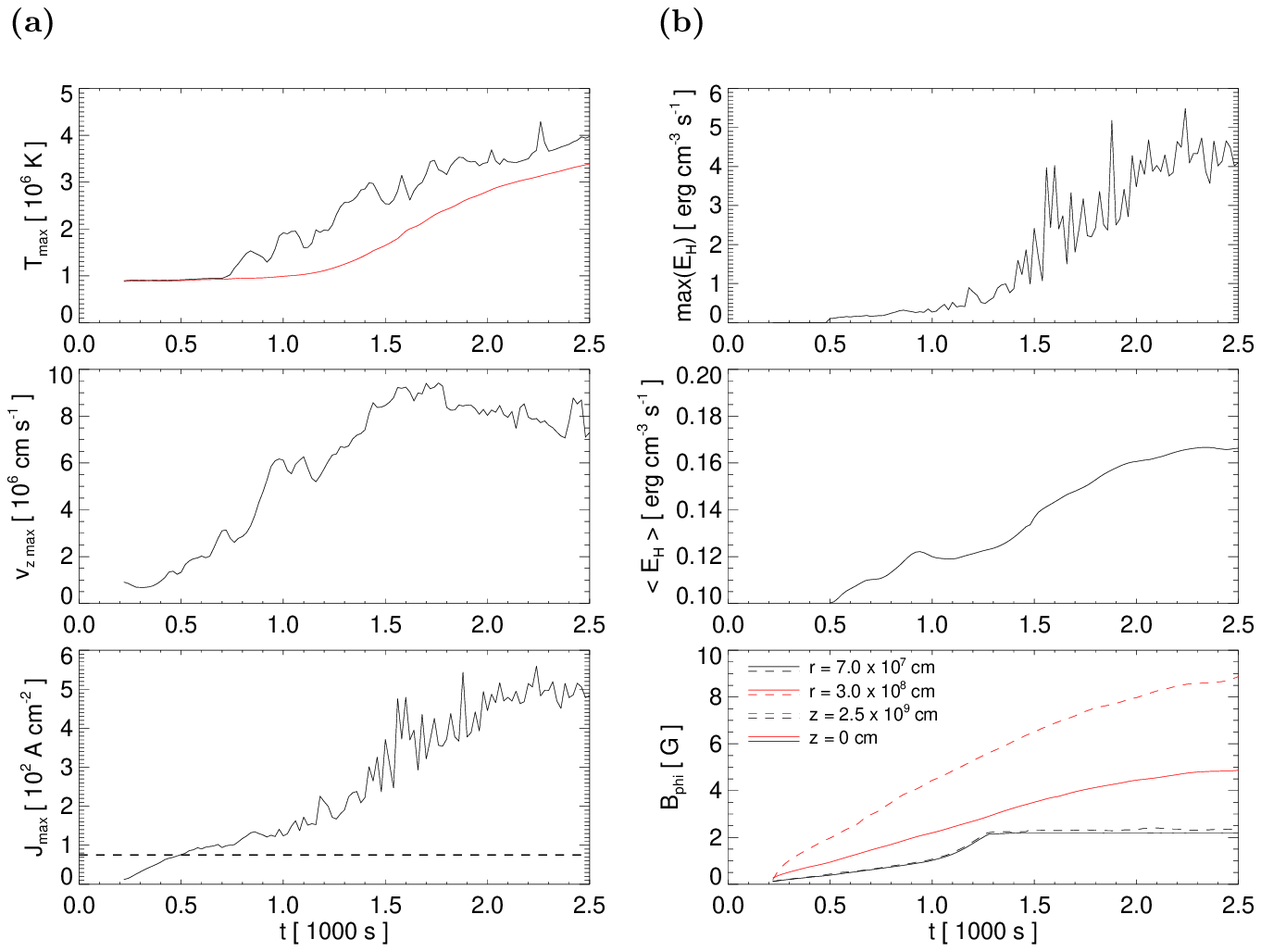}}
\caption{\small Evolution of (a) the maximum temperature, vertical velocity and coronal maximum current density in the simulated box, and (b) of the maximum heating rate per unit volume (top), the averaged heating over cells with  $E_H > 0$ and $r < 10^9$ cm (middle, black solid lines) and \aftr{of the azimuthal component of the magnetic field $B_{\phi}$, for $\phi = 45^o$, at the two labelled heights $z$ along the loop, i.e., apex (solid) and just above the transition region (dashed) and at the two labelled radial distances $r$ from the central axis, i.e., close to the axis (black) and 3000 km far away (red).} In panel (a) the average temperature at the loop apex (red line) and current threshold for dissipation (horizontal dashed line) are also shown.   }
\label{fig:evol1d}
\end{figure}

Fig.~\ref{fig:evol1d}a shows the evolution of the maximum loop temperature, maximum vertical speed and maximum current density in the region where the dissipation is allowed, i.e. above the transition region. The maximum current density has an increasing trend until $t \sim 2000$ s, although with strong fluctuations at late times. Then it seems to become steady. The threshold for dissipation is reached at $t = t_h \approx 500$ s and at the end of the simulation the maximum value is above 500 A cm$^{-2}$. Fig.~\ref{fig:prof}b shows that these high values are localized in the low corona. 

The maximum temperature is initially steady at $\sim 1$ MK and it begins to increase with an irregular trend at time $t \approx 700$ s, taking about $t \sim 1000$ s to settle around the maximum of $\sim 4$ MK, a hot active region loop. In spite of the slight delay, the overall temperature trend resembles quite closely the one of the maximum current. The top panel of Fig.~\ref{fig:evol1d}a shows also the average temperature at the loop apex, which gives an idea of the average conditions of the loop. This temperature begins to rise for $t > 1000$ s and reaches a value above 3 MK at the end time, typical of active region loops.
From Fig.~\ref{fig:evol1d}a we see that the maximum temperature does not rise as long as the resistivity is off. Therefore, the effect of the enhanced magnetic tension due to the twisting is low, at least in the corona.

The maximum vertical speed provides information about the strength of the evaporation. It begins to increase readily at $t \approx t_h$ and takes about 1000 s to settle to $\sim 80 - 100$ km/s, with a slightly decreasing trend for $t > 1500$ s. These relatively high values of speed are typical of impulsive evaporation, driven by a continuous sequence of heat pulses \citep[e.g.,][]{Patsourakos2006a}. 

Fig.~\ref{fig:evol1d}b shows the evolution of the maximum heating rate and of the heating rate averaged only over the heated cells, i.e., with $E_H > 0$ and $r < 10^9$ cm.  The evolution of the maximum heating rate resembles closely that of the maximum current density (Fig.~\ref{fig:evol1d}a), with spikes reaching very high values ($\sim 5$ erg cm$^{-3}$ s$^{-1}$). Each spike represents an impulsive energy release. The duration of each pulse is less than a minute, and these are the high energy tails of a distribution that provides the average heating rate produced by the twisting and shown in the bottom panel of Fig.~\ref{fig:evol1d}b. This average rate increases by about 60\% throughout the simulation. \aftr{In the last panel, Fig.~\ref{fig:evol1d}b shows the evolution of the azimuthal component of the magnetic field $B_{\phi}$, i.e., of the twisting, at two different heights along the loop (apex and just above the transition region) and at two different radial distances from the central loop axis (close to the axis and 3000 km apart). At first $B_{\phi}$ invariably increases at all position, more rapidly far from the axis because the rotation is faster there. At time $t \sim 1300$ s it saturates close to the loop axis, because of the dissipation. Farther from the axis the curves saturate much later, close to the end of the simulation. There, $B_{\phi}$ grows less at the apex than at the bottom, because of the field expansion, i.e. the field is weaker at the top than below. Close to the axis, instead, the expansion is small and the field component grows more uniformly.}

\begin{figure}[!ht]               
\centering
  {\includegraphics[width=4cm]{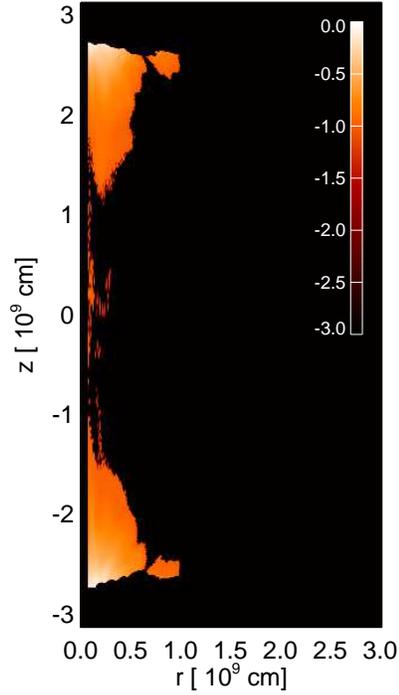}}
\caption{\small Cross-section in the r-z plane across the loop central axis of the volumetric heating rate, $E_{H}$, at t = 2500 s (log scale). 
}
\label{fig:heating}
\end{figure}

\begin{figure}[!ht]               
\centering
  {\includegraphics[width=12cm]{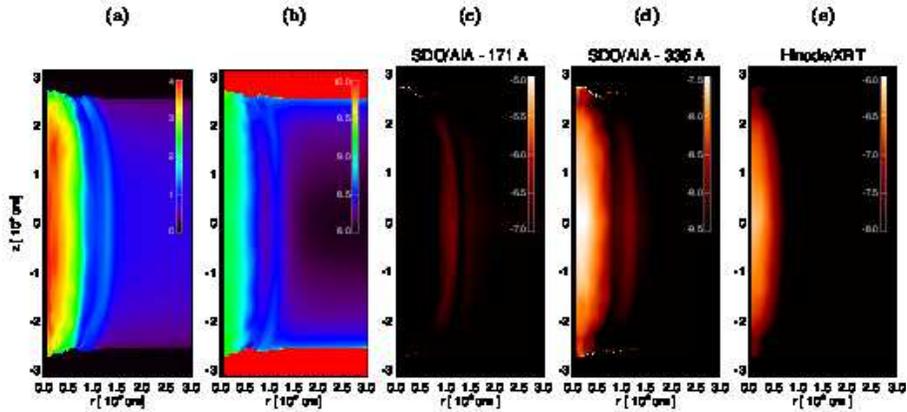}}
\caption{\small From left to right: Cross sections of the temperature and density, and of the synthetic emission (log scale)  in the SDO/AIA 171 {\AA} and 335 {\AA} channels and in the Hinode/XRT Ti\_poly filter band, at t = 2500 s. The emission units are DN cm$^{-1}$ s$^{-1}$ pix$^{-1}$.}
\label{fig:emimage}
\end{figure}

\new{Fig.~\ref{fig:heating} shows information about the distribution of the heating release at the final time, which can be compared to the current distributions shown in Figs.~\ref{fig:evol3d_jcur} and \ref{fig:prof}. The figure shows the cross-section of the heating distribution across the central axis in the $r-z$ plane. The heating is clearly broader and more intense near the loop footpoints, some minor quantity is released along the central axis. Some heating is released also for $r > 5 \times 10^8$ cm, only close to the footpoints.}
%
Fig.~\ref{fig:emimage} shows the temperature and density and the emission predicted from a slice across the loop central axis in two EUV (SDO/AIA 171 \AA\ and 335 \AA) channels and in one X-ray (Hinode/XRT Ti\_poly) channel. In the 171 \AA\ channel, sensitive to plasma at $\sim 1$ MK, the loop is practically invisible. We only see a faint halo in the outer shell and bright layers at the footpoints. This is expected because the loop plasma is mostly at temperatures around 2-3 MK, and therefore only the thin transition region emits in this channel. The 335 \AA\ channel is more sensitive to plasma hotter than 2 MK and the loop is fully visible and bright in this channel. For the same reason it is analogously bright in the X-ray band: here only the central region is bright because this channel is more sensitive to higher temperature plasma. The emission predicted in these two channels is fairly uniform in the body of the loop, and the loop appears as monolithic. Some inhomogeneity and tapering is present close to the footpoints.

\subsection{Comparison with constant resistivity and uniform rotation}

\begin{figure}[!ht]               
\centering
  {\includegraphics[width=10cm]{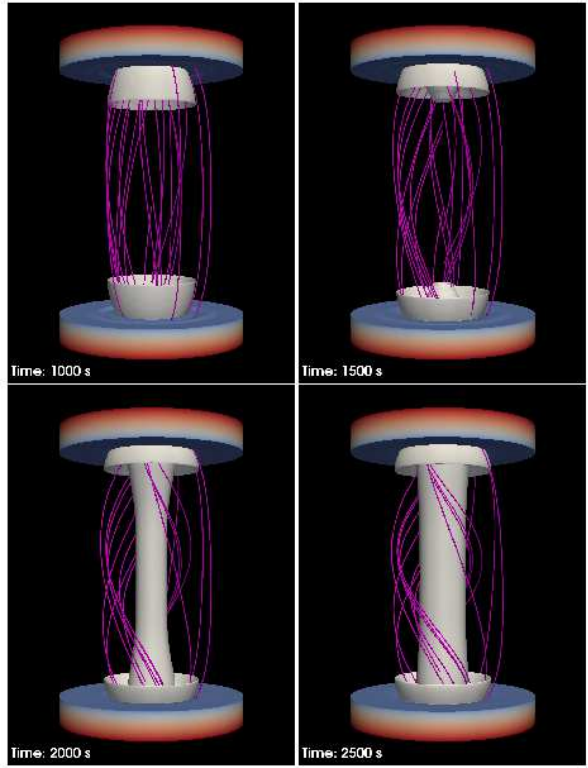}}
\caption{\small Current density for the simulation with constant coronal resistivity and unpertubed footpoint rotation to be compared with Fig.~\ref{fig:evol3d_jcur}. }
\label{fig:evol3d_jcur_runi}
\end{figure}

\begin{figure}[!ht]               
\centering
  \subfigure[]
 {\includegraphics[width=6cm]{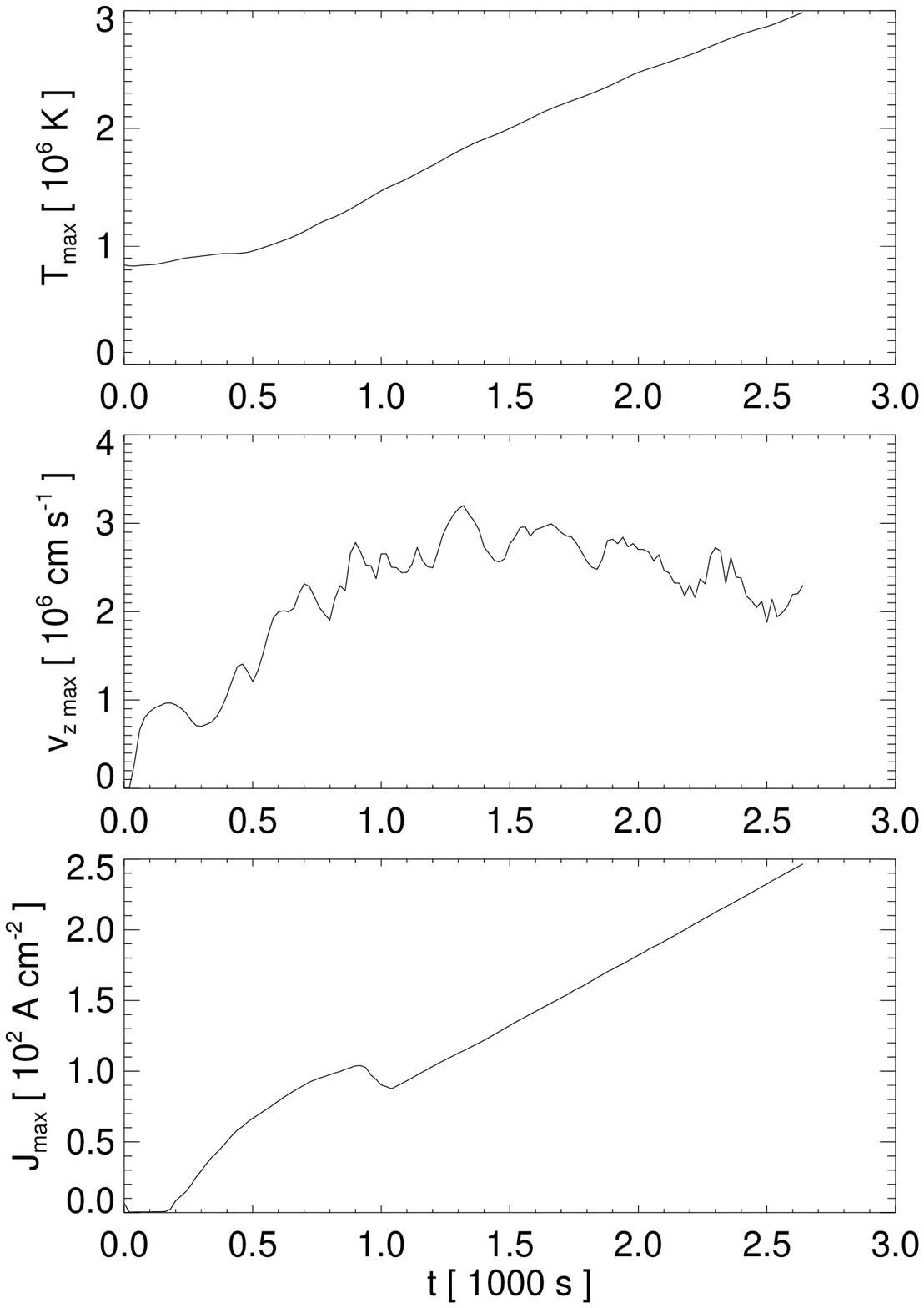}}
\caption{\small As Fig.~\ref{fig:evol1d}a, for the simulation with constant coronal resistivity and unpertubed footpoint rotation. }
\label{fig:evol1d_runi}
\end{figure}

To understand the role of the selected heated mechanism, we have compared the reference simulation above with another identical one (hereafter CRS) except for two issues: a) the resistivity is constant and always on in the corona (see Section~\ref{sec:resist}), with no current threshold, and a value $\eta = 10^{13}$  cm$^2$ s$^{-1}$, i.e. 10 times lower than the switch-on value (Section~\ref{sec:resist}); b) the velocity field at the footpoints is not random, i.e. there is a uniform rotation motion.  This choice implies a radial symmetry around the central vertical axis, and we actually obtain a radially symmetric evolution. According to this simulation, the flux tube is gradually heated to coronal temperature and filled with plasma from the chromosphere. All this evolution occurs more gradually and uniformly than in the reference simulation. Fig.~\ref{fig:evol3d_jcur_runi} shows that the current density grows from the loop footpoints upwards and eventually it is high uniformly all along the loop axis. The fine structure that we see in Fig.~\ref{fig:evol3d_jcur} is purely due to the perturbations of the rotation motion at the footpoints, that are not present in the simulation of Fig.\ref{fig:evol3d_jcur_runi}. Fig.\ref{fig:evol1d_runi} shows other interesting features from the comparison with the reference case  (Fig.\ref{fig:evol1d}). The temperature regime is analogous, so the comparison is sound, but the reference case shows higher peaks, above 4 MK. The reference simulation also yields much higher evaporation speeds and currents (more than twice as high on average). This difference is due to the presence of the threshold for current dissipation, which lets the magnetic field be stressed more and energy be released more impulsively.

\section{Discussion and conclusions}

\new{This work describes a possible scenario of a coronal loop heated by magnetic field stressing. This is an evolution from the standard one-dimensional single, or multi-strand, loop modeling (see Section~\ref{sec:intro}) and a step forward in self-consistent MHD loop modeling.} Single loop models describe the hydrodynamics of a coronal atmosphere linked to the chromosphere through a steep transition region confined in a curved flux tube. The curvature appears only in the formulation of the gravity. The plasma moves and transports energy only along the tube, under the effect of a prescribed heating function. Here we maintain the same plasma atmosphere but we immerse it in an ambient ``cylindrical" magnetic field, which expands from the chromosphere up into the corona, as in \cite{Guarrasi2014a}. We no longer consider a prescribed heating function, but the heating is a consequence of stressing the magnetic field. This provides a self-consistent conversion of magnetic energy into heat. Our choice has been to start from simple, but realistic, assumptions on magnetic stressing and energy conversion: the magnetic field is stressed through the twisting driven by rotational footpoint motion in layers where plasma $\beta >> 1$; twisting is believed to be quite usual in the solar atmosphere \citep{{Wedemeyer-Bohm2012a},De-Pontieu2014a}. The rotation motion is perturbed as expected in the solar surface, and this is fundamental to break the symmetries and let currents fragment into sheets \citep{Rappazzo2013a,Nickeler2013a}. The sheets are progressively intensified by the twisting and this occurs more at the loop footpoints where the magnetic field is tapered crossing the transition region to the chromosphere. In this scenario we have hypothesised a switch-on dissipation mechanism. Heating from a very high anomalous resistivity is released as soon as the current density grows above a given threshold \citep{Hood2009b}, that we set to $2.25 \times 10^{11}$ esu s$^{-1}$ cm$^{-2}$, to mimic possible turbulent cascades or MHD avalanche \citep{Rappazzo2013b,Hood2016a}. Since here we address mostly the coronal evolution, the heating is assumed (in common with some previous simulations) to be active in the corona only, just because otherwise the magnetic field in the chromosphere is rapidly dissipated and we have found no way to refurbish it.

Our single loop study supports other findings from MHD modeling of solar atmosphere boxes \citep[e.g.][]{Hansteen2015a}, and provides fine details. We start from a tenuous and cool atmosphere. Where the current grows above the threshold in the corona, the plasma begins to heat above 1 MK. The heating is more steady and efficient around the loop central axis, where the temperature rises above 3 MK on average in about half an hour. At the same time, the increasing pressure gradients determine the expansion of the chromospheric layers and the tube fills with denser plasma. The density gradually rises above $10^9$ cm$^{-3}$. This evaporation is in agreement with standard single loop models. From comparison with an equivalent simulation with ever-present anomalous resistivity, we have ascertained that the presence of the switch-on heating that leads to a factor two larger evaporation speeds, even in the late steady state. This is therefore a major difference between a gradual and an impulsive heating mechanism. Another important difference is the presence of overheated plasma. At variance from the gradual-heating simulation, the switch-on heating produces some amount of plasma significantly hotter than the average, as expected from impulsive heating \citep{Klimchuk2006a} and recently detected in bright active regions \citep[e.g.][]{Reale2009b,Reale2011a,Testa2012c,{Miceli2012a}}.

The heating is more intense where the magnetic field is more intense, i.e. close to the footpoints, where it expands more. This is in agreement with other MHD modeling of the solar atmosphere \citep{Gudiksen2005a,Gudiksen2005b,Bingert2011a,Bingert2013a}. The heating is more intense around the central axis where the footpoint rotates and the twisting of the magnetic field is effective. The energy release determines a progressive dissipation of the magnetic field, through the local reconnection of sheared field lines. With our choice of magnetic diffusivity, this dissipation drives the plasma to density and temperature typical of active regions already at moderate twisting angles, far from the conditions to trigger kink instabilities \citep{Hood1979b,Hood1981a,Einaudi1983a,Velli1990a,Torok2003a}.

The kink instability has been suggested as a trigger mechanism for the rapid heating of coronal loops. \cite{Hood2009b} have shown that its non linear development creates current sheets, and triggers magnetic reconnection. Once reconnection starts, the current sheets fragment, resulting in the dissipation of magnetic energy across the loop cross-section, as it relaxes towards its lowest energy state. Starting from a temperature of only $10^4$~K, their results show that plasma heating up to $10^7$~K and above is possible. \cite{Botha2011a} included thermal conduction so that lower temperatures were obtained.

In addition to previous twisting models, our description includes the chromosphere and the transition region at reasonable resolution. Another important ingredient is the expansion of the magnetic field from the chromosphere to the corona \citep{Rosner1978b}, which, together with the change of $\beta$ regime in the chromosphere, stresses the importance of the non-linear interaction of the plasma and the magnetic field. Our model is also able to describe a significant mass transfer from the chromosphere to the corona, an essential feature for comparison with the observed loops brightness. 
The plasma produces realistic emission in X-ray and EUV bands. Its evaporation and the twisting drive significant spiraling motions as recently extensively observed \citep{De-Pontieu2014a}. 

Our choice here is to produce loop heating with a relatively ordered magnetic stressing, i.e. the progressive random twisting due to footpoint rotation. So we are not describing an entirely chaotic magnetic stress, determined by random photospheric motions that lead to magnetic braiding \citep{Lopez-Fuentes2010a,Wilmot-Smith2011a,Bingert2011a}. Our approach allows us to keep a tighter grasp on the physical effects that lead to the loop evolution, still maintaining a reasonable description of possible coronal drivers \citep{Rosner1978b}.

An entirely ordered footpoint rotation would not lead to the formation of structured currents and heating. As mentioned above, an essential ingredient to have fine structure is a random motion at the footpoints. The deriving fine structure is both in space and time. We see filamentary structures on the cross-scale of a few hundreds kilometers, but also a structured heating with spikes that reach the scale of proper flare intensities, with durations on the scale of few tens of seconds. Evidence for loop fine structure is widespread \citep[e.g.][]{Vekstein2009a,Guarrasi2010a,{Viall2011a},{Antolin2012a},{Brooks2012b},{Brooks2013a},{Cirtain2013a},{Peter2013a},{Testa2013a},Tajfirouze2016a,{Tajfirouze2016b}}. 

The fine temporal and spatial structure that develops within our modeling deserves further investigation and will be the subject of future research. In particular, we plan to study the effects of: radial motions; different rotation profiles; different  magnetic field strengths; and different initial magnetic configurations, especially those that lead to tectonics heating \citep{Priest2002a}.

\acknowledgements{FR, SO, GP acknowledge support from italian Ministero dell'Universit\`a e Ricerca. We acknowledge PRACE for awarding us access to resource FERMI based in Italy at CINECA through the project no. 2011050755 ``The way to heating the solar corona: finely-resolved twisting of magnetic loops''. FR thanks ISSI Bern for the support to the team "Coronal Heating - Using observables (flows and emission measure) to settle the question of steady vs. impulsive Heating". PLUTO is developed at the Turin Astronomical Observatory in collaboration with the Department of Physics of the Turin University.  }

\bibliographystyle{apj}

\end{document}